\documentclass[aps,prd,notitlepage,nofootinbib,superscriptaddress,twocolumn]{revtex4-1}

\usepackage[utf8]{inputenc}
\usepackage{amsmath}
\usepackage{amssymb}
\usepackage{amsfonts}
\usepackage{newtxtext,newtxmath}
\usepackage{bm}
\usepackage{graphicx}
\usepackage[usenames,dvipsnames]{xcolor}
\usepackage{color}
\usepackage[colorlinks=true,linkcolor=blue,urlcolor=blue,citecolor=blue]{hyperref}

\usepackage{slashed}
\usepackage[english]{babel}
\usepackage{dcolumn}
\usepackage{blindtext}
\usepackage{epsfig}
\usepackage{pifont}
\usepackage{dsfont,mathrsfs}
\usepackage{cancel}
\usepackage{bigints}
\usepackage{accents}
\usepackage{soul}
\usepackage{multirow}
\usepackage[normalem]{ulem}

\usepackage{soul}
\usepackage{simpler-wick}
\usepackage{float}

\def\beq{\begin{equation}}
\def\eeq{\end{equation}}
\def\beqa{\begin{eqnarray}}
\def\eeqa{\end{eqnarray}}
\def\ban{\begin{eqnarray*}}
\def\ean{\end{eqnarray*}}
\def\bi{\begin{itemize}}
\def\ei{\end{itemize}}

\begin{document}

\title{Quark matter under strong electric fields in the Linear Sigma Model coupled with quarks}

\author{William R. Tavares} \email{wr.tavares@yahoo.com}
\affiliation{Departamento de Física Teórica, Universidade do Estado do Rio de Janeiro, 20550-013 Rio de Janeiro, RJ, Brazil}  

\author{Sidney S. Avancini}\email{sidney.avancini@ufsc.br}
\affiliation{Departamento de F\'{\i}sica, Universidade Federal de Santa
  Catarina, 88040-900 Florian\'{o}polis, SC, Brazil} 
\author{ Ricardo L. S. Farias}\email{ ricardo.farias@ufsm.br}
\affiliation{Departamento de F\'{\i}sica, Universidade Federal de Santa
  Maria, 97105-900 Santa Maria, RS, Brazil}

\begin{abstract}
 In this work we study the influence of external electric field and temperature on the chiral phase transition of Quantum Chromodynamics. We use the two-flavor Linear Sigma Model coupled with quarks (LSMq) in a thermal and electrized medium to evaluate the effective quark mass and the Schwinger pair production. To this end, we apply one-loop correction to the fermionic sector of the model and the simple tree-level approximation in the mesonic contributions. The electric fields strengthen the partial restoration of the chiral symmetry when applied with finite temperature in a crossover transition. The expected decrease of the pseudocritical temperature as a function of the electric field is observed until electric fields reach $eE\approx 13.5 m_{\pi}^2$. For stronger electric fields, the effect is the opposite, which is in a very good agreement with previous results obtained with four-point non-renormalizable models, showing that this effect is independent of renormalizability issues. We also show the thermal and electric effects on the behavior of the Schwinger pair production.
\end{abstract}

\maketitle

\section{Introduction}

Experimental efforts over the last few decades can give us very strong evidence of the formation of quark-gluon plasma in the relativistic heavy-ion collisions (HIC) \cite{Busza:2018rrf,Elfner:2022iae,Snellings:2011sz}. If one assumes basic ideas from classical electromagnetism, one can obtain that strong magnetic fields perpendicular to the reaction plane can be present in peripheral heavy-ion collisions with $eB\sim10^{19}$ G\cite{Huang:2015oca}. This is a preliminary sketch of more sophisticated computational techniques that predict magnetic fields of the same order of magnitude depending on the impact parameter, electrical conductivity and the collision time \cite{Tuchin:2013apa,Tuchin:2015oka,McLerran:2013hla}. Besides that, recent numerical simulations predict that strong magnetic and electric fields should be present in peripheral HIC, which is the case observed in asymmetrical collisions \cite{Deng:2014uja,Hirono:2012rt,Voronyuk:2014rna,Cheng:2019qsn}. In such situations, e.g., Cu$+$Au collisions, strong electric fields are expected due to the difference of electric charges in the region where the nuclei overlap. It is also feasible to anticipate the emergence of strong electric fields by predictions of event-by-event fluctuations of proton position within the colliding nuclei of Au+Au and Pb+Pb \cite{Bzdak:2011yy,Deng:2012pc,Bloczynski:2012en,Bloczynski:2013mca} at the usual collision energy scale $\sqrt{s}=2.76$ TeV and $\sqrt{s}=200$ GeV from ALICE and RHIC respectively. Additionally, the presence of electric fields can be very interesting to better understand anomalous transport properties as the chiral separation effect \cite{Huang:2015oca,Huang:2013iia,Jiang:2014ura,Pu:2014fva}, the chiral magnetic effect \cite{Fukushima:2008xe}  and several other quantities in the phase diagram of quantum chromodynamics (QCD).

Despite the numerical evidence about strong electric fields in HIC, there is still little effort to increment QCD phase diagram analysis in such an environment. In lattice QCD  (LQCD), the main reason concerns technical issues similar to the sign problem \cite{Yamamoto:2012bd,Yamamoto:2021oys} which makes very difficult such applications, regardless of a few recent works with some improvements \cite{Endrodi:2021qxz,Endrodi:2022wym,Yamamoto:2021oys,Yang:2022zob}. In this way, the current literature about electric fields in QCD is restricted basically to low energy effective models and some applications in quantum field theories (QFTs), such as the case of Nambu--Jona-Lasinio (NJL) model and its extensions \cite{Tavares:2018poq,Tavares:2019mvq,Cao:2015dya,Ruggieri:2016xww,Ruggieri:2016lrn,Cao:2015dya,Klevansky:1989vi,Klevansky:1992qe}, Chiral Perturbation Theory \cite{Cohen:2007bt,Tiburzi:2008ma}, Dyson-Schwinger equations \cite{Ahmad:2020ifp} and $\lambda\phi^4$ theory \cite{Loewe:2021mdo,loewe2022effective,Loewe:2022aaw}. Most of these models are in good agreement with regard to the partial restoration of the chiral symmetry guided by electric fields, namely inverse electric catalysis (IEC). In the case of the NJL model, some different regularization techniques have also been used to evaluate not just the behavior of the effective quark mass but also the Schwinger pair production \cite{Cao:2015dya}, even in more complex environments including electric fields \cite{Ruggieri:2016lrn,Ruggieri:2016xww}. If simultaneously to the electric fields we include the temperature, the two-flavor NJL \cite{Tavares:2018poq,Tavares:2019mvq} results indicate that both quantities will strengthen the partial chiral symmetry restoration, with the decreasing of the pseudocritical temperature as a function of the electric fields until $eE\sim 13.5m_{\pi}^2$, where an opposite behavior is predicted, which until now is due inconclusive reasons \cite{Tavares:2019mvq}. The same qualitative result has been obtained in the context of the $\lambda\phi^4$ self-interacting scalar field theory \cite{Loewe:2021mdo}.

In the present work we explore for the first time, the two-flavor Linear Sigma model (LSM$q$) coupled with quarks with a constant electric field and finite temperatures. We apply the one-loop correction in the fermionic sector of the model and the usual tree-level approximation for the mesonic sector. Inspired by the regularized expressions of the pure electric field part in the thermodynamic potential, developed in the context of NJL model \cite{Tavares:2018poq,Tavares:2019mvq}, we apply these set of equations in the renormalizable version of LSM$q$. Our aim is to reanalyze all the basic physics of the model, i.e., effective quark masses, Schwinger pair production, and observe how the pseudocritical temperature for chiral symmetry restoration behaves as a function of the electric fields. These quantities must be enough to know how the renormalizability affects the results in comparison with previous results obtained in NJL model.

The work is organized in the following structure: In section \ref{sec2} we present the formalism details of LSM$q$ including effects of an electric field and finite temperature. In section \ref{sec3} we show the equations of the one-loop correction for the fermionic contribution including thermo-electric effects. The numerical results are present in section \ref{sec4} and the conclusions in section \ref{sec5}. The appendix~\ref{appA} is devoted to the explicit computation of the minimum of the effective potential.

\section{Lagrangian of the SU(2) LSM$_q$ with electric fields}\label{sec2}

The Lagrangian of the SU(2) LSM$_q$ in an external electromagnetic field is given in Euclidean space by the following expression \cite{Andersen:2014xxa}

\begin{eqnarray}
\mathcal{L}&=&\overline{\psi}\left[i \slashed D +g(\sigma + i\gamma_5\vec{\tau}\cdot\vec{\pi})\right]\psi+\frac{1}{2}\left[(\partial_{\mu}\sigma)^2+(\partial_{\mu}\vec{\pi})^2\right]\nonumber\\
           && + U(\sigma,\vec{\pi})-\frac{1}{4}F_{\mu\nu}F_{\mu\nu,}\label{lag}
\end{eqnarray}

\noindent where $A_\mu$ and $F_{\mu\nu} = \partial_\mu A_\nu - \partial_\nu A_\mu$ 
are respectively the electromagnetic gauge and  tensor fields, $g$ is the Yukawa coupling constant, $\vec{\tau}$ are isospin Pauli matrices,  $Q$ is the diagonal quark charge 
\footnote{Our results are expressed in Gaussian natural units 
where $1\,{\rm GeV}^2= 1.44 \times 10^{19} \, G$ and $e=1/\sqrt{137}$.} matrix, 
$Q$=diag($q_u$= $2 /3$, $q_d$=-$1/3$), 
$D_\mu =(\partial_{\mu} + i e Q A_{\mu})$ is the covariant derivative and we adopt $A_{\mu}=-\delta_{\mu 4}x_3E$ in order to include in the $z-$direction a constant electric field. In this model, we have the following quantum fields:   
 $\psi = (\psi_u \quad \psi_d)^T$ which are the quark fermion fields and the mesonic degrees of freedom as given by the $\sigma$ and $\vec{\pi}$ fields. The purely mesonic potential, $U(\sigma,\vec{\pi})$, in eq.(\ref{lag}), is given by
 \begin{eqnarray}
  U(\sigma,\vec{\pi}) = \frac{1}{2}m^2\left(\sigma^2+\vec{\pi}^2\right)+\frac{\lambda}{24}\left(\sigma^2+\vec{\pi}^2\right)^2 - h\sigma
 \end{eqnarray}

 \noindent where $\lambda, m^2$ and $h$ are constants fixed by experimental parameters as the $\sigma$ and $\pi$ meson masses and the pion decay constant, $f_{\pi}$. The value of $h=f_{\pi}m_{\pi}^2$ ensure the explicit breaking of the chiral symmetry, i.e., $SU(2)_v$ group. We adopt, for simplicity the mean field values of $\langle \sigma \rangle \equiv \phi$ and $\langle \vec{\pi}\rangle\equiv \vec{\pi}=0$.

The tree-level effective potential in a constant electric field for the LSM$_q$ is given by

\begin{eqnarray}
 F^{0}=U(\phi,\vec{\pi}) - 
 \frac{1}{4}E^2,\label{omega0}
\end{eqnarray}

\noindent this representation is useful for making clear some of our renormalization procedures to the one-loop fermionic correction to the effective potential.
 
The meson and quark masses, at the tree-level approximation, are given by \cite{Andersen:2014xxa}

\begin{eqnarray}
 && m_{\sigma}^2=m^2+\frac{\lambda}{2}\phi^2,\\
 && m_{\pi}^2=m^2+\frac{\lambda}{6}\phi^2,\\
 && M = g\phi
\end{eqnarray}

\noindent where $\phi$ is vacuum expectation value of $\sigma$ field.

\section{One-loop fermionic contribution}\label{sec3}

In this work we treat mesonic sector at tree level and include quantum corrections only in the fermionic sector. The one-loop correction to the fermionic contributions at $eE=T=0$ is given by \cite{Skokov:2010sf,Ayala_2021,Ayala_2018,Ayala:2017ucc}

\begin{eqnarray}
 F^{1}_{vac}=-2N_cN_f\int\frac{d^4p}{(2\pi)^4}\log(p_0^2+E_p^2) + C,\label{vacfermion}
\end{eqnarray}

\noindent where $C$ is a mass-independent constant that can be ignored, $N_c$ and $N_f$ are the number of colors and flavors, respectively. The energy dispersion relation, $E_p$, is given by

\begin{eqnarray}
 E_p^2=M^2+p^2,
\end{eqnarray}

In the $\overline{\text{MS}}$ scheme \cite{Andersen:2014xxa}, we obtain for eq.(\ref{vacfermion}) the following expression

\begin{eqnarray}
  F^{1}_{vac}=\frac{N_cN_fM^4}{16\pi^2}\left[\log\left(\frac{\Lambda^2}{M^2}\right)+\frac{3}{2}\right].\label{vacuumfermion}
\end{eqnarray}
\noindent where $\Lambda$ is the renormalization scale.
In the present scheme, one needs to include the counter-terms $m^2\rightarrow m^2+\Delta m^2$ and $\lambda\rightarrow \lambda+\Delta \lambda$ \cite{Andersen:2014xxa}, where

\begin{eqnarray}
 &&\Delta m^2 = \frac{\lambda m^2}{16\pi^2\epsilon},\\
 &&\Delta \lambda = \frac{2\lambda^2-24N_cN_fg^4}{16\pi^2\epsilon}\\
 &&\delta \varepsilon = \frac{m^4}{16\pi^2\epsilon},
\end{eqnarray}

\noindent where we have included a vacuum energy counter-term $\Delta\varepsilon$ \cite{Andersen:2014xxa}.

When considering a constant electric field, the one-loop fermion contribution can be written in the proper-time formalism \cite{Schwinger:1951nm}

\begin{eqnarray}
 F^{1}=\sum_f\frac{N_c}{8\pi^2}\int_0^{\infty}ds\frac{e^{-sM^2}}{s^2}\mathcal{E}_f\cot(\mathcal{E}_fs),\label{omegaqq}
\end{eqnarray}

\noindent where $\mathcal{E}_f=|q_feE|$. We can verify the validity of the last expression just evaluating the analytical extension derived by Schwinger \cite{Schwinger:1951nm}, i.e., $eE\rightarrow ieB$, where we obtain the exact expression to the LSM$_q$ with a constant magnetic field in the $z-$ direction \cite{Tavares:2018poq,Tavares:2019mvq}. 

The integration in eq.(\ref{omegaqq}) has divergences with different sources. First, we treat the divergence in the lower limit of integration, $s= 0$, by using the separation of divergences from Schwinger's work in QED \cite{Schwinger:1951nm}. For this purpose we use the Taylor expansion of $\cot(\mathcal{E}_fs)$ function for $\mathcal{E}_fs\ll1$

\begin{eqnarray}
 \cot(\mathcal{E}_fs)\sim \frac{1}{\mathcal{E}_fs}-\frac{\mathcal{E}_fs}{3}-\frac{(\mathcal{E}_fs)^3}{45}+\mathcal{O}((\mathcal{E}_fs)^5),\quad \mathcal{E}_fs\ll1\nonumber.
\end{eqnarray}

 Using the expansion in eq.(\ref{omegaqq}), we can avoid the divergences in the region $s=0$

\begin{eqnarray}
 F^{1}(\mathcal{E},M)&=&\sum_f\frac{N_c}{8\pi^2}\int_0^{\infty}ds\frac{e^{-sM^2}}{s^3}\left(\mathcal{E}_fs\cot(\mathcal{E}_fs)-1\right.\nonumber\\
 &+&\left.\frac{\left(\mathcal{E}_fs\right)^2}{3}\right)
 +\frac{N_c}{8\pi^2}\sum_f\int_{\frac{1}{\Lambda^2}}^\infty ds\frac{e^{-sM^2}}{s^3}
 \nonumber\\
 &-&\frac{N_c}{24\pi^2}\sum_f\mathcal{E}_f^2\int_{\frac{1}{\Lambda^2}}^\infty ds\frac{e^{-sM^2}}{s}.
 \label{F1}
\end{eqnarray} 
 
 In eq.(\ref{F1}) the first three terms are the finite electric field contributions, which were regularized with respect to the ultraviolet divergence from the contributions near $s=0$. In the second term of the second line, we have the fermionic contribution from the vacuum given in eq. (\ref{vacuumfermion})  and the field contribution, proportional to $\mathcal{E}^2$, that must be regularized and renormalized. For simplicity, we define the one-loop fermionic contribution as

\begin{eqnarray}
     F^{1}(\mathcal{E},M)&=& F^{1}_{vac}(M)+F^{1}_{med}(\mathcal{E},M)+F^{1}_{field}(\mathcal{E},M),\nonumber\label{omegaqq1}
\end{eqnarray}
 
 \noindent where we have separated $F^1$ in the vacuum, medium and field contributions. Each term is given by the following expressions:

 \begin{eqnarray}
   F^{1}_{med}(\mathcal{E},M)&=&  \sum_f\frac{N_c}{8\pi^2}\int_0^{\infty}ds\frac{e^{-sM^2}}{s^3}\nonumber\\ &&\times\left(\mathcal{E}_fs\cot(\mathcal{E}_fs)-1+\frac{(\mathcal{E}_fs)^2}{3}\right),\\
   F^{1}_{vac}(M)&=& \frac{N_c}{8\pi^2}\sum_f\int_{\frac{1}{\Lambda^2}}^\infty ds\frac{e^{-sM^2}}{s^3},\nonumber\\
   &=&\frac{N_cN_fM^4}{16\pi^2}\left[\log\left(\frac{\Lambda^2}{M^2}\right)+\frac{3}{2}\right],\\
   F^{1}_{field}(\mathcal{E},M)&=&-\frac{N_c}{24\pi^2}\sum_f\mathcal{E}_f^2\int_{\frac{1}{\Lambda^2}}^\infty ds\frac{e^{-sM^2}}{s}\nonumber\\
   &=& -\frac{N_c}{24\pi^2}\sum_f\mathcal{E}_f^2\left[-\log\left(\frac{M^2}{\Lambda^2}\right)-\gamma_E\right],
 \end{eqnarray}
 
 \noindent where $\gamma_E$ is the Euler-Mascheroni constant.
 Usually, the divergence proportional do $E^2$ is renormalized and incorporated in the $\frac{1}{2}E^2$ from the Lagrangian eq.(\ref{lag}). For renormalization purposes, we set $E^2\rightarrow Z^2E^2$, where:
 
 \begin{eqnarray}
  Z^2=\left[1+N_c\sum_f\frac{4q_f^2}{3(4\pi)^2\epsilon}\right].
 \end{eqnarray}

The details of the last contribution are evaluated in Ref. \cite{Andersen:2014xxa}, in the case of a constant magnetic field, which in our case we do not consider the one-loop mesonic contributions.

Now we work with the divergent contributions present in eq.(\ref{omegaqq1}), that occurs in the proper-time integration when $\mathcal{E}_fs=n\pi$ in the cotangent function, with $n=0,1,2,3...$. These divergences are associated with the instabilities in the vacuum due to the electric field, which gives rise to the Schwinger pair production of quarks. We extend analytically to the complex plane the integration and separate the real from the imaginary part. This procedure is developed in detail in Ref.\cite{Tavares:2018poq}. This gives rise to the following result

\begin{eqnarray}
 \Re\left(F^{1}_{med}(\mathcal{E},M)\right)&=&\frac{N_c}{2\pi^2}(\mathcal{E}_f)^2 
\left \{ \zeta'(-1) + \frac{\pi}{4}y_f + \frac{y_f^2}{2} \left( \gamma_E \right.\right.\nonumber \\
&&-\frac{3}{2}+\left.\left.\ln y_f \right) - \frac{1}{12} \left(1+ \ln y_f \right)  \right. \nonumber\\ 
&&\left.+\sum_{k=1}^{\infty}k\left[\frac{y_f}{k}\tan^{-1}\left(\frac{y_f}{k}\right) \right.\right.\nonumber\\
&&- \frac{1}{2}\ln\left(1+\left(\frac{y_f}{k}\right)^2\right)-\left.\left.\frac{1}{2}\left(\frac{y_f}{k}\right)^2\right] \right \}, 
\label{potef1}
\end{eqnarray}
 
 \noindent where $y_f=\frac{M^2}{2\mathcal{E}_f}$ and $\zeta'(-1) = 1/2-\log(A) $ with $A=1.282417...$ being the Gleisher-Kinkelin constant \cite{Gleisher}. The imaginary part of the one-loop effective potential is
 \begin{eqnarray}
  \Im\left(F^{1}_{med}(\mathcal{E},M,T)\right)=\frac{N_c}{4\pi}\sum_{f}\mathcal{E}_f^2\sum_{k=1}^{\infty}\frac{e^{-\frac{M^2\pi k}{\mathcal{E}_f} }}{(k\pi)^2}\label{decay}.
 \end{eqnarray}

The thermo-electric contribution is evaluated in the imaginary-time formalism. The result is given by

\begin{eqnarray}
 \Re\left(F^{1}_{therm}(\mathcal{E},M,T)\right)&=&- \frac{N_c}{2\pi^2}\sum_{n=1}^{\infty}(-1)^n\mathcal{E}_f\int_0^{\infty}ds\frac{e^{-sM^2_f}}{s}\nonumber\\
 &&\times \cot(\mathcal{E}_fs)e^{-\frac{\mathcal{E}_fn^2}{4|\tan(\mathcal{E}_fs)|T^2}}. \label{phi001}
\end{eqnarray}
where the summation is over the Matsubara frequencies. The thermo-electric contribution is finite due to the Fermi-Dirac distribution, written in the proper-time formalism. The full thermodynamic potential, including the tree-level potential $F^0$ in eq.(\ref{omega0}), is given, therefore, by

\begin{align}
    \Re\left(F(\mathcal{E},M,T)\right) = \Re\left(F^0+F^{1}(\mathcal{E},M) + F^{1}_{therm}(\mathcal{E},M,T)\right).
\end{align}

It is possible to show that, given the thermodynamic potential, we can derive the Schwinger pair production rate, i.e., $\Gamma(\mathcal{E},T,M)=\Im\left(F^{1}(\mathcal{E},M,T)\right)$ \cite{Tavares:2018poq,Tavares:2019mvq}. 

In order to fix $\Lambda$, at $\mathcal{E}=T=0$, we choose the physical point $\phi=f_{\pi}$ at the minimum of the effective potential \cite{Andersen:2014xxa}, i.e.,
\begin{eqnarray}
    \frac{d(\Re F(\mathcal{E},M,T))}{d\phi}|_{\phi=f_{\pi}}=0,\label{fixL}
\end{eqnarray}

To obtain the behavior of the effective quark mass as function a of the temperature and electric fields, we evaluate the gap equation, given in Appendix \ref{appA}.

\section{Numerical Results}\label{sec4}

In this section we present and discuss our numerical results. The model requires fixing three independent parameters: the mass parameter $m$, the boson-fermion coupling $g$ and the boson self-coupling $\lambda$. These parameters are fixed in such a way to obtain with LSMq the vacuum values of pion mass,  $m_{\pi}=140$ MeV,  pion decay constant, $f_\pi=93$ MeV,
the $\sigma$ meson mass, $m_{\sigma}=800$ MeV and effective quark masses $M=300$ MeV.
With these experimental inputs, we obtain the following set of parameters:  $g=3.2258$, $m^2=-290600$ MeV$^2$  and $\lambda=215.19$. We also need to fix the renormalization scale $\Lambda$, having determined the parameters $m$, $g$ and $\lambda$ we can choose the value of $\Lambda$ where the minimum of the one loop effective potential in the vacuum remains the same at tree level value ($\phi = 93$ MeV). Similar procedure was made in~\cite{Andersen:2014xxa} and the value obtained for the renormalization scale was $\Lambda=181.96$ MeV.

\begin{figure}[h]
\begin{tabular}{ccc}
\includegraphics[width=8.5cm]{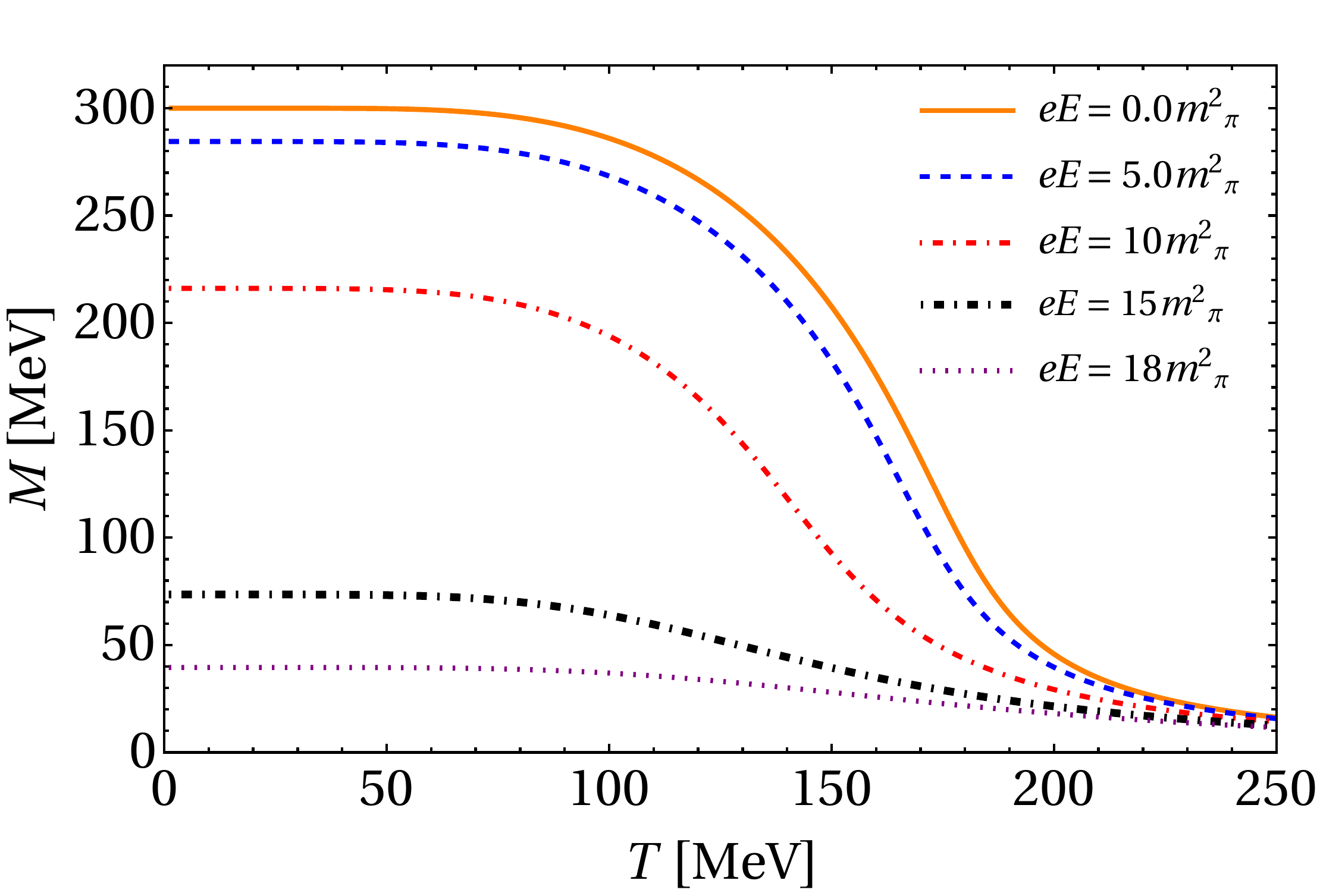}\\
\end{tabular}
\caption{Effective quark mass as a function of the temperature for different values of electric fields.}
\label{MxT}
\end{figure}

In Fig.~\ref{MxT} we can see the effective quark mass as a function of the temperature for different values of electric fields. For low values of electric fields, we can see the partial restoration of chiral symmetry as we increase the temperature. This effect is strengthened for higher values of electric fields, which is very clear for strong enough values, e.g., $eE=15m_{\pi}^2$ and $eE=18m_{\pi}^2$. Additionally, for all of these cases, the transition is a crossover. Despite that the two effects combine themselves to partially restore the chiral symmetry, they are different physical phenomena, i.e., the temperature excites the system weakening  the interaction in the chiral condensate. On the other hand, the electric fields accelerate the charges with different signs in opposite directions, inducing the chiral condensate to be less probable to happen as we increase the strength of the electric field.

When we increase the electric fields, the transition occurs to lower values of temperatures, indicating inverse electric catalysis (IEC) of the pseudocritical temperature. However, in the purple dotted line, for $eE=18m_{\pi}^2$, we cannot distinguish by the Fig.~\ref{MxT} in a very clear way if we have electric catalysis (EC) or IEC. The phenomena EC and IEC will be easier to verify when we plot the pseudocritical temperature as a function of the electric field.

\begin{figure}[h]
\begin{tabular}{ccc}
\includegraphics[width=8.5cm]{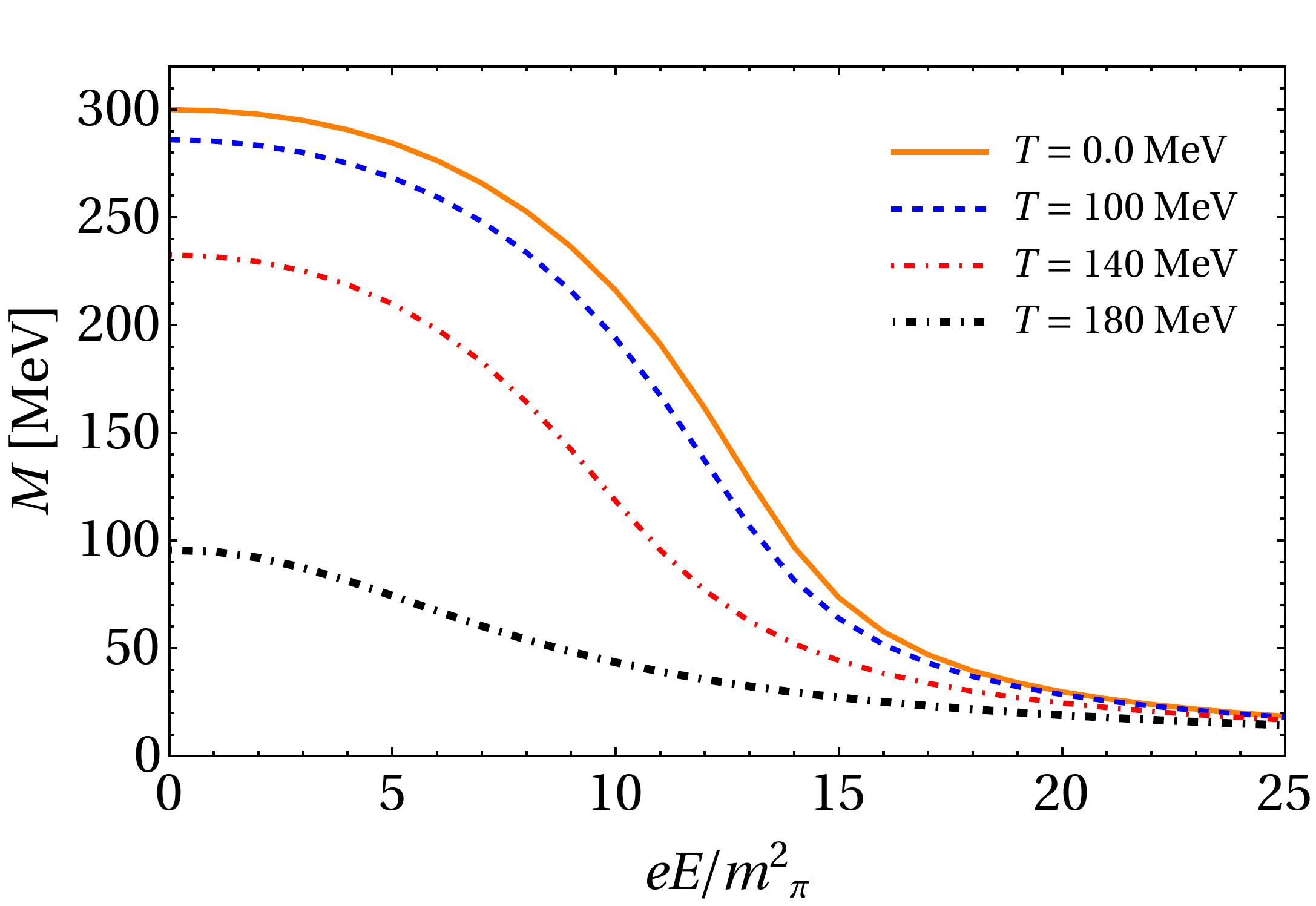}\\
\end{tabular}
\caption{Effective quark mass as a function of the electric field for different values of temperature.}
\label{MxeE}
\end{figure}

We can see the effective quark masses as a function of the electric fields for different values of temperature in Fig.~\ref{MxeE}. The analysis indicates that, at low temperatures, the electric fields partially restore the chiral symmetry breaking at $eE\gg15m_{\pi}^2$ as indicated previously in Fig.~\ref{MxT}. For low values of the electric fields, we can see the influence of the thermal effects on the partial restoration of the chiral symmetry. 
On the other hand, as we increase the strength of the electric field, Fig.2 shows the effect of the electric field on the partial restoration of the chiral symmetry.

\begin{figure}[h]
\begin{tabular}{ccc}
\includegraphics[width=8.5cm]{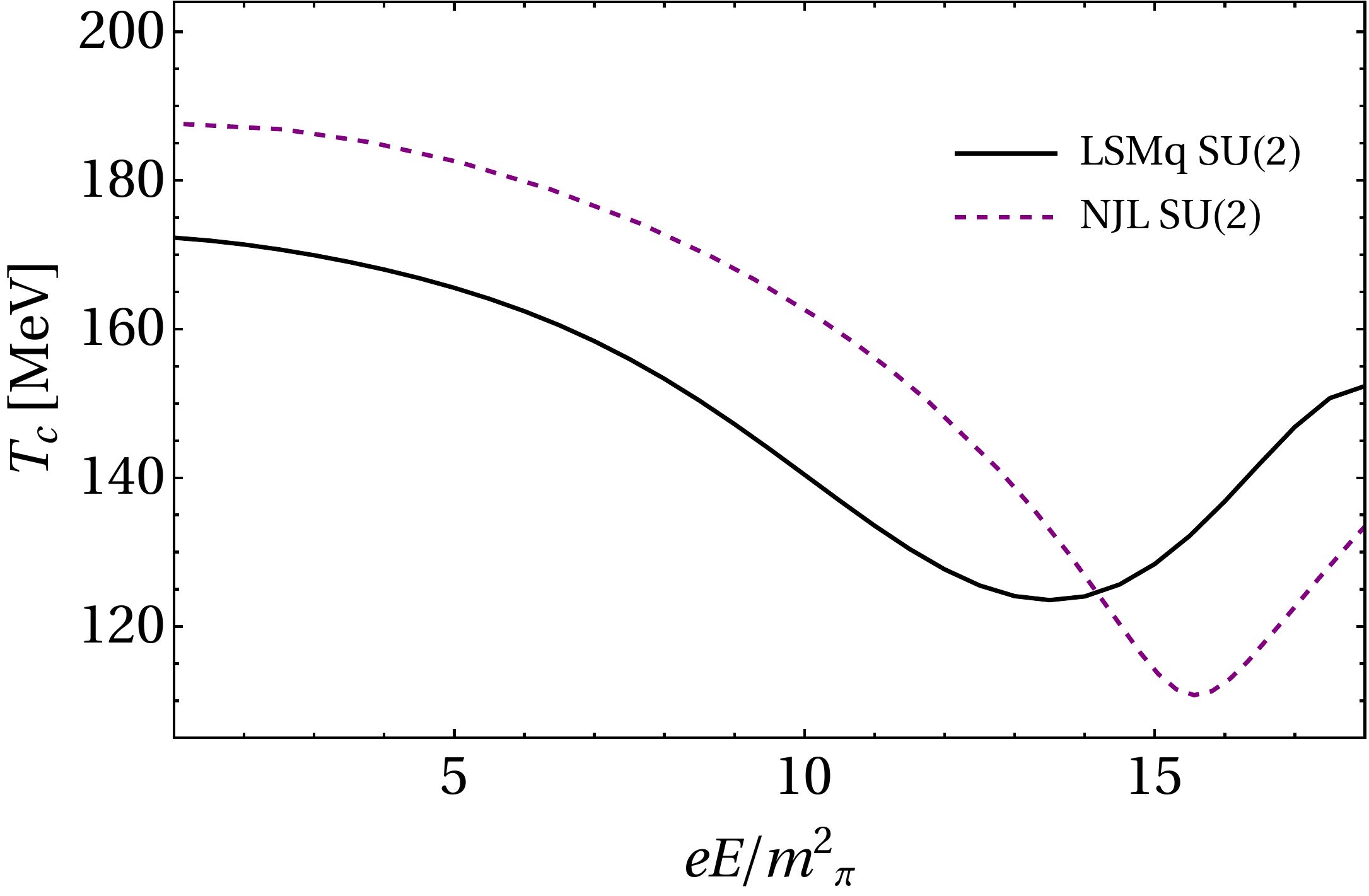}\\
\end{tabular}
\caption{The pseudocritical temperature for chiral symmetry restoration as a function of the electric field in the LSMq (solid line) and NJL SU(2) (dashed line) \cite{Tavares:2019mvq}.}
\label{TcxeE}
\end{figure}

In the Fig.~\ref{TcxeE}, we show the pseudocritical temperature for chiral symmetry restoration as a function of the electric field. 
The pseudocritical temperature is calculated as the maximum of the $- \partial M/\partial T$.
We can clearly see the IEC effect, which turns into EC for very high values of electric fields, i.e., $eE>13.5m_{\pi}^2$, where we have the increasing of the pseudocritical temperature as a function of electric fields. This result is in good qualitative agreement with the previous works with SU(2) NJL and Polyakov--Nambu--Jona-Lasinio (PNJL) models~\cite{Tavares:2019mvq}. 
Therefore, the phenomenon observed previously in both NJL and PNJL that the IEC change to EC after a critical value of the electric field can not be attributed to the regularization issues, which are inherent to nonrenormalizable models. Since our results show now by using a renormalizable model qualitatively the same behavior which has been observed before in NJL and PNJL models.

\begin{figure}[h]
\begin{tabular}{ccc}
\includegraphics[width=8.5cm]{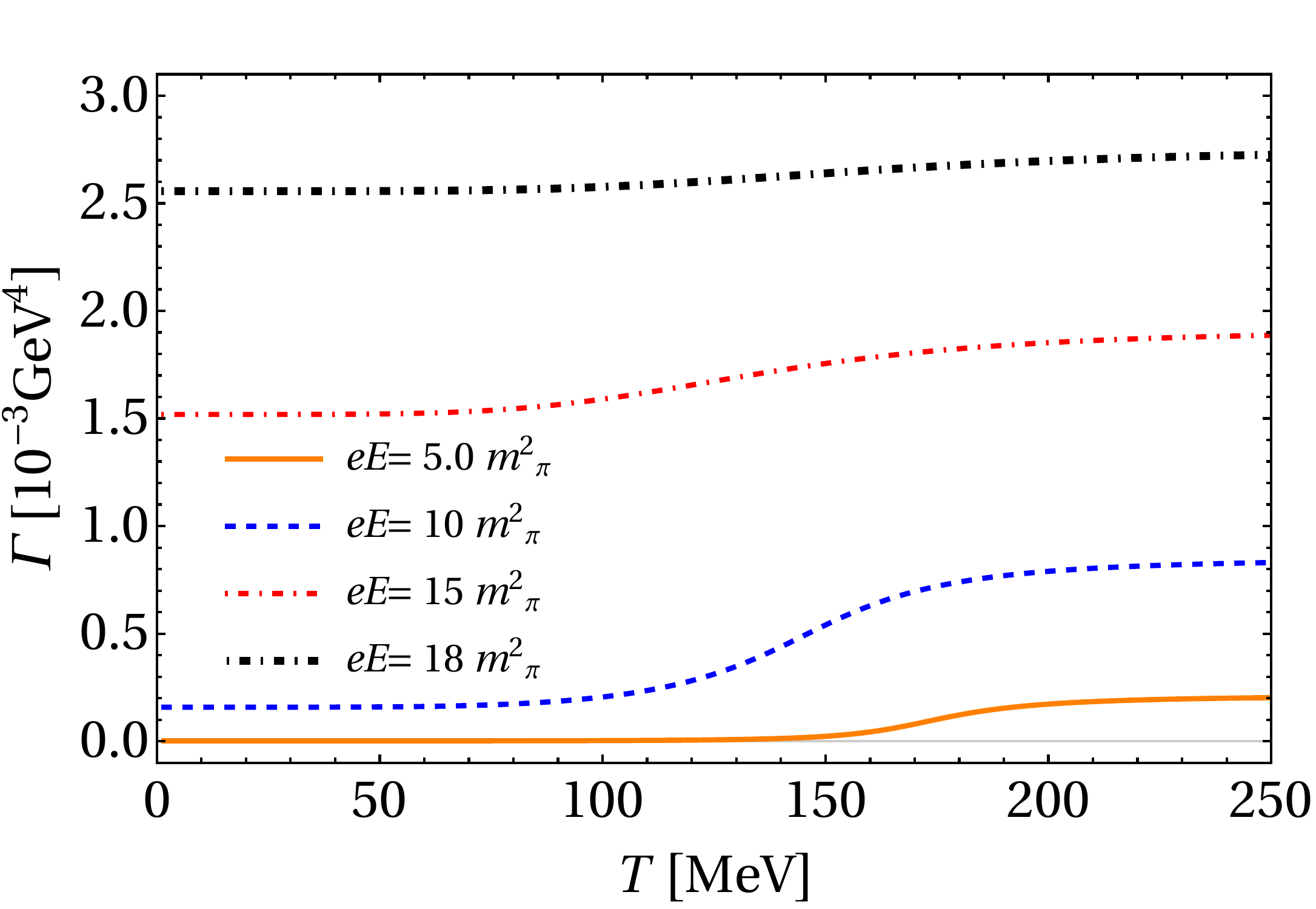}\\
\end{tabular}
\caption{The Schwinger pair production of quarks as a function of the temperature for different values of the electric fields.}
\label{PPxT}
\end{figure}

The Schwinger pair production as a function of temperature is given in Fig.~\ref{PPxT}. We can see that, for lower values of electric fields, e.g., $eE=5m_{\pi}^2$, the production is almost zero in the low-temperature limit,  and slightly changes at $T>T_{pc}$, indicating that the high temperatures can strengthen the pair production in an electric medium. As we increase the electric fields, the pair production grows in the low-temperature region and substantially increases after the pseudocritical temperature, which we can see in the situations $eE=10m_{\pi}^2$ and $eE=15m_{\pi}^2$. At very high electric fields, the Schwinger pair production almost does not change in the full temperature range considered. These results are in good agreement with the previous two-flavor NJL model results~\cite{Tavares:2018poq,Tavares:2019mvq}. 

\begin{figure}[h]
\begin{tabular}{ccc}
\includegraphics[width=8.5cm]{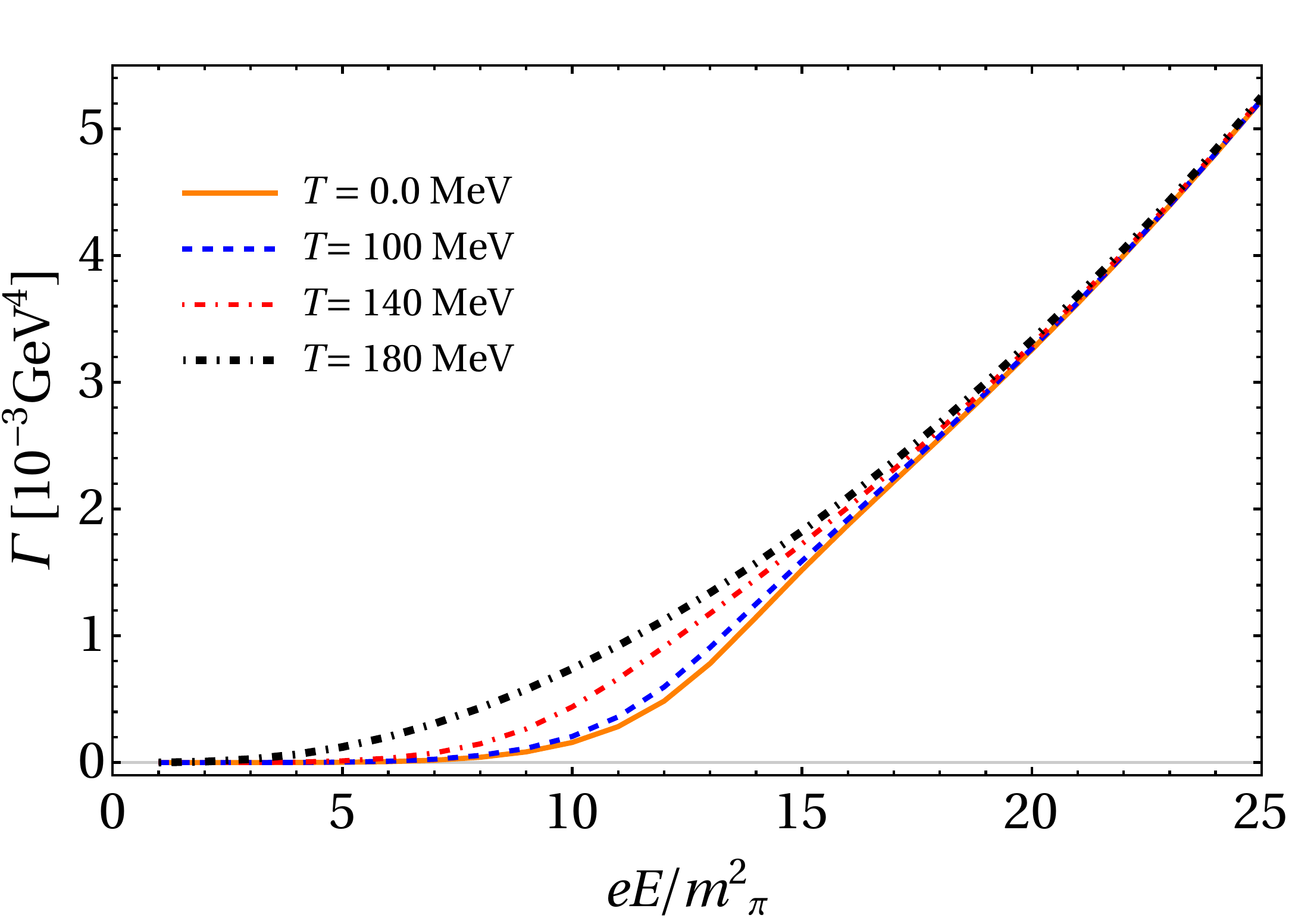}\\
\end{tabular}
\caption{The Schwinger pair production of quarks as a function of the electric fields for different values of the temperature.}
\label{PPxeE}
\end{figure}

In Fig.~\ref{PPxeE} are shown our results for the Schwinger pair production as a function of the electric field normalized by the pion mass squared. It is clear from this figure that non-negligible effects 
start to appear around $eE = 5 m_\pi^2$ and for fixed electric fields the temperature enhances the pair production. Above $eE=18m_\pi^2$ the pair production saturates. Again, these results are in qualitative agreement with previous NJL and PNJL calculations. 

\section{Conclusions}\label{sec5}

In this work, we have studied within the Linear sigma model with quarks, the quark matter at finite temperature with a background of electric fields. One of the main motivations of this study was the estimation of the effects of the strong electric fields on the chiral symmetry restoration scenario within a renormalizable model, and performing the comparison with previous results obtained in SU(2) versions of NJL and PNJL  models~\cite{Tavares:2019mvq}.

Our numerical results have shown that the constituent quark masses decrease when we increase the strength of the electric fields, as a signature of the partial restoration of the chiral symmetry. 
We have computed the evolution of the pseudocritical temperature for chiral symmetry restoration $T_{pc}$ as a function of the electric fields. In the literature, as expected, usually the pseudocritical temperature for chiral symmetry restoration decreases as we increase the electric field strength. In~\cite{Tavares:2019mvq} within SU(2) PNJL model we have shown, for the first time in the literature, a very interesting effect where for strong enough electric fields the pseudocritical temperature for chiral symmetry restoration starts to increase after a critical value of the electric field. This effect was founded also within LSMq and also propagates to all quantities, as to the Schwinger pair production. In the context of $\lambda\phi^4$ theory similar results were found, which show IEC for weak electric fields and EC for strong electric field strength \cite{Loewe:2021mdo}. Our work provides strong evidence that the non-monotonic behavior of the pseudocritical temperature of chiral symmetry restoration as a function of the electric field is a characteristic of QCD, regardless of whether we are working with a renormalizable or non-renormalizable effective model of QCD.

An improvement of the present work is the study of the inclusion of baryon density effects in the presence of thermal and background electric fields. We are currently exploring these avenues in PNJL and LSMq and will report on the findings elsewhere in the near future.

\appendix

\section{Gap equation}\label{appA}

In this section, we will deal with the derivative of the real part of the effective potential. For simplicity, we will define $\Re (F)\equiv F$.
The minimum of the effective potential is given by, $ dF/d\phi =0$,  where $M=g\phi$, as following

\begin{eqnarray}
   &&\frac{dF^0}{d\phi}+\frac{dF^1_{vac}(M)}{d\phi}+\frac{dF^{1}_{med}(\mathcal{E},M)}{d\phi}+\nonumber\\ &&\frac{dF^{1}_{field}(\mathcal{E},M)}{d\phi}+\frac{d F^{1}_{therm}(\mathcal{E},M,T)}{d\phi}=0,\label{derivpot}
\end{eqnarray}

\noindent where the first term in Eq.(\ref{derivpot}) concerns to the tree-level potential, which gives
\begin{eqnarray}
\frac{dF^0}{d\phi} &=& \frac{dU(\phi,\vec{\pi})}{d\phi}\nonumber\\
&=& m^2\phi + \frac{\lambda}{6}[\phi^2+\vec{\pi}^2]\phi - h\nonumber\\
&=& m^2\phi + \frac{\lambda}{6}\phi^3 - h\nonumber
\end{eqnarray}

\noindent where we assume that the mean field value $\vec{\pi}\rightarrow\langle\vec{\pi} \rangle=0$. The second term in Eq.(\ref{derivpot}) is the one-loop correction to the fermionic contribution, given by

\begin{align}
   \frac{dF^{1}_{vac}(M)}{d\phi}=\frac{N_cN_fgM^3}{4\pi^2}\left[\log\left(\frac{\Lambda^2}{M^2}\right)+\frac{3}{2}\right]-\frac{N_cN_fgM^3}{8\pi^2}.
\end{align}

The medium and field contributions combine themselves as

\begin{eqnarray}
&&\frac{d F^{1}_{med}(\mathcal{E},M)}{d\phi}+\frac{d F^{1}_{field}(\mathcal{E},M)}{d\phi}\nonumber\\&=&-\sum_f\frac{g M N_c}{4\pi^2}\int_0^{\infty}ds
\frac{e^{-sM^2}}{s^2}\left[\mathcal{E}_fs\cot\left(\mathcal{E}_fs\right)-1\right]   \nonumber \\ 
&&=\sum_f\frac{gMN_c}{2\pi^2}\mathcal{E}_f \left[ \frac{\pi}{4}+y_{f}(\gamma_E-1+\ln y_{f})+\right. \nonumber \\
&&+\left. \sum_{k=1}^{\infty}\left(\tan^{-1}\frac{y_{f}}{k}-\frac{y_{f}}{k} \right)\right]. \label{reg1}
\end{eqnarray}

The last term in eq.(\ref{derivpot}) concerns the one-loop thermoelectrical contribution, which is given by

\begin{eqnarray}
    \frac{d F^{1}_{therm}(\mathcal{E},M,T)}{d\phi}&=&-\frac{gMN_c}{2\pi^2}\sum_{n=1}^\infty(-1)^n\int_0^{\infty}ds\frac{e^{-sM^2}}{s}\mathcal{E}_f\nonumber\\&&\times\cot(\mathcal{E}_fs)
  e^{-\frac{\mathcal{E}_fn^2}{4|\tan(\mathcal{E}_fs)|T^2}}.\label{thetaET0}
\end{eqnarray}

Therefore, we obtain the gap equation for numerical analysis.

\section*{Acknowledgments}

This work was partially supported by Conselho Nacional de Desenvolvimento Cient\'ifico 
e Tecno\-l\'o\-gico  (CNPq), 309598/2020-6 (R.L.S.F.), 
304518/2019-0 (S.S.A.), 309262/2019-4; Fundação Carlos Chagas Filho de Amparo à Pesquisa do 
Estado do Rio de Janeiro (FAPERJ), Grant No.SEI-260003/019544/2022 (W.R.T); 
Funda\c{c}\~ao de Amparo \`a Pesquisa do Estado do Rio Grande do Sul (FAPERGS), Grants Nos. 19/2551- 0000690-0 
and 19/2551-0001948-3 (R.L.S.F.); The work is also part of the 
project Instituto Nacional de Ci\^encia e Tecnologia - F\'isica Nuclear e Aplica\c{c}\~oes 
(INCT - FNA), Grant No. 464898/2014-5. 

\bibliographystyle{spphys}
\bibliography{refs.bib}
\end{document}